\documentclass[12pt]{iopart}

\usepackage{iopams}
\usepackage[dvips]{graphicx}
\usepackage{booktabs}
\usepackage{scrtime} 

\begin{document}

\title{Daylight operation of a free space, entanglement-based quantum key
  distribution system}

\author{Matthew P. Peloso$^{1,2}$, Ilja Gerhardt$^{1}$, Caleb Ho$^{1}$, Ant\'{\i}a Lamas-Linares$^{1,2}$ and Christian Kurtsiefer$^{1,2}$}

\address{$^1$ Centre for Quantum Technologies, National University of Singapore, 3 Science Drive 2, Singapore, 117543}
\address{$^2$ Department of Physics, National University of Singapore, 2 Science Drive 3, Singapore, 117542}

\ead{christian.kurtsiefer@gmail.com}

\begin{abstract}

\noindent Many quantum key distribution (QKD) implementations using a free
space transmission path are restricted to operation at night time in order to
distinguish the signal photons used for a secure key establishment from
background light. Here, we present a lean entanglement-based QKD
system overcoming that limitation. By implementing spectral, spatial and
temporal filtering techniques, we were able to establish a secure key
continuously over several days under varying light and weather conditions.
\end{abstract}


\maketitle

\section{Introduction\label{intro}}
Since its inception by Bennett and Brassard in 1984
\cite{bennett:84}, quantum cryptography has made the transition from a concept
to a technology mature enough for commercial development
\cite{magiq, idquantique}. There are several flavors of quantum
cryptography or quantum key distribution. The initial formulation, and
all current commercial systems implement so-called prepare and send (PaS)
protocols
\cite{scarani:08}, where some degree of freedom of light is prepared by
one party, Alice, and sent to the other party, Bob, who then measures it
in one out of several complementary bases.
Estimation of the errors in the measurement results of the
receiver allows both parties to place an upper bound on the knowledge of an
eavesdropper, and is used for a subsequent removal of this knowledge
in a privacy amplification step.

Another family of protocols evolved out of a proposal by Ekert in 1991 (E91,
\cite{ekert:91}). These protocols use entanglement as the main resource, and
combine some of the measurements on the biphotons such that a Bell inequality
can be tested, or the state is tomographically estimated \cite{enzer:02} to
evaluate the knowledge of an eavesdropper. 
 An important development was an explicit way to calculate the amount of information leaked to an eavesdropper out of a
less than perfect violation of a Bell inequality \cite{acin:07}, which makes it possible
to implement this idea in a practical system with imperfect sources and measurement devices \cite{ling:08}. These protocols reduce the
assumptions about the physical implementation (like
e.g. the size of the Hilbert space used to encode information) in comparison
with most PaS QKD schemes.

An entanglement-based BB84-type QKD scheme was described by
Bennett, Brassard and Mermin in 1992 (BBM92, \cite{bennett:92}). There, the
prepare part of BB84 is replaced by a measurement scheme similar to the
receiver side, but the knowledge of an eavesdropper is still evaluated from
the observed errors. This results in a larger fraction of final key
bits than under a full E91 protocol in its quantitative version
\cite{acin:07}, and probably maintains the
insensitivity against an unknown size of the Hilbert space, as long as the
measurement devices can be trusted. Furthermore, this scheme retrieves the
randomness for the
key used for encryption directly out of the measurement process on a quantum
system, and does not need to provide for an active choice of a key bit. This
QKD scheme has been demonstrated in
the field \cite{tittel:98, poppe:04} using optical fiber links without
amplifiers or signal regeneration stages. If the link is to be established
\emph{ad hoc}, e.g. in a mobile environment, or it is
not feasible to have a fiber deployed (e.g. in the satellite QKD proposals
\cite{hughes:99, ursin:08}), propagation of the photons
through free space is necessary.
A free space transmission channel using polarization encoding of the
qubits has the advantage of not inducing decoherence (negligible
birefringence of air), and has low absorption under clear weather conditions.

So far, such entanglement based QKD systems over free space have been
demonstrated at night, taking advantage of low background light levels
\cite{ling:08, peng:05, marcikic:06, ursin:07,  erven:08}. In this paper we
demonstrate
daylight operation of a QKD system implementing a BBM92 protocol. Continuous
operation over a full day/night cycle brings free space entanglement based QKD
one step closer to the stage of development of free space PaS protocols, where 
such daylight operation has been shown \cite{hughes:00, hughes:02}.

\section{Background rate estimations} \label{chap:theory}
The main challenge for operating over a free space channel in daylight is to
handle the high background from the sun. First, actively quenched avalanche
photodiode (APD) detectors may be subject to irreversible
destruction when exposed to an excessive amount of light; such a situation may
occur if there is excessive scattering in the optical communication link.
For passively quenched APDs this is not a problem, since the electrical power
deposited into the device can be limited to a safe operation regime at all
times.

Second, saturation of detectors leads to a reduced probability of detecting
photons at high light levels. This effect can usually be modeled by a dead
time $\tau_{\rm d}$ or recovery time for the device. For passively quenched
APDs, this time is about $1\,\mu$s, but may be over  an order of magnitude
smaller for actively quenched devices.
While modeling the saturation with a single dead time $\tau_{\rm d}$ may not
completely reflect the details of the re-arming of a detector, it gives a useful
estimation of the fraction of time a detector can register photoevents. Given
an initial photoevent rate $r$ (i.e., the rate a detector with no recovery
time would report), a detector with dead time $\tau_{\rm d}$ will register a rate of

\begin{equation}\label{eq:saturation}
  r'=r(1-r'\tau_{\rm d})\quad\textrm{or}\quad r'=r{1\over1+r\tau_{\rm d}}.
\end{equation}

Third, a high background level will lead to detection events which are
mistaken with the detection of a photon
pair. These are  uncorrelated in their polarization and lead to an
increase in the quantum bit error ratio (QBER), which is used to establish a
bound for the knowledge of an eavesdropper. In the following, we
estimate the operational limit for generating a useful key under such
conditions, assuming an implementation of a symmetrical BBM92 protocol, i.e.,
both complementary measurement bases are chosen with an
equal probability of 50\% on both measurement units.

Assuming that all quoted rates already include detector efficiencies, we
can characterize a pair source by its single event rates, $r_1, r_2$, and its
coincidence rate $r_{\rm c}$. We denote the transmission of the entire
optical channel as $T$, in which we include absorptive losses in optical
components, the air, geometrical losses due to imperfect mode transfer from an
optical fiber, and losses in spatial filters.

The signal or raw key rate for a symmetric BBM92 protocol is given by half of
the detected coincidence rate,
\begin{equation}
r_{\rm sig}=\frac{1}{2}r_{\rm c}T\,.
\end{equation}

\noindent For an external background event rate $r_{\rm bg}$, a
coincidence time interval of $\tau_{\rm c}$, and assuming no correlations
between source and background events, the accidental coincidence rate with
matching bases is given by 
\begin{equation}
r_a= \frac{1}{2} (r_1-Tr_{\rm c})\left(r_{\rm bg}+ 
  T(r_2-r_{\rm c})\right)\tau_{\rm c}\,,
\end{equation}

\noindent assuming that only one of the detectors, here with index 2, is
exposed to the background events.

Imperfections in practical entangled photon pair source and the
detector projection errors are often characterized by visibilities
of polarization correlations $V_{HV}$ and $V_{\pm45^\circ}$. The intrinsic
QBER $q_{\rm i}$ of the QKD system with 
a symmetric usage of both bases is given by

\begin{equation}
  q_{\rm i}={1\over2}\left(1-{V_{HV}+V_{\pm45^\circ}\over2}\right)\,.
\end{equation}

\noindent The polarization of background events on one side can be assumed to
be uncorrelated the photons detected in the other arm, thus the QBER due to
accidentally identified coincidences is $1/2$. The total QBER $q_{\rm t}$ of
the complete ensemble is given by the weighted average over both components,
\begin{eqnarray} \label{eq:qbertot}
  q_{\rm t}&=&{1\over r_{\rm sig}+r_a}\left(q_{\rm i}r_{\rm sig}+
    {1\over2}r_a\right) \nonumber \\
&=&
  {q_{\rm i}r_{\rm c}T+(r_1-Tr_{\rm c})\left(r_{\rm bg}+T(r_2-r_{\rm
        c})\right)\tau_{\rm c}/2\over 
    r_{\rm c}T+(r_1-Tr_{\rm c})\left(r_{\rm bg}+T(r_2-r_{\rm
        c})\right)\tau_{\rm c}}\,.
\end{eqnarray}

\noindent The detector saturation modifies both signal and accidental rates
similarly to equation (\ref{eq:saturation}) by the same dead time correction factor
$\alpha$, where we assume an equal distribution of photoevents over all four
detectors, resulting in a dead time constant of $\tau_{\rm d}/4$:

\begin{equation}
\alpha={1\over1+(r_{\rm bg}+r_2T)\tau_{\rm d}/4}\,.
\end{equation}

\noindent Therefore, the resulting QBER $q_{\rm t}$ in equation (\ref{eq:qbertot}) does
not get affected. However, the signal rate does, leading to the modified
expression

\begin{equation}
  r'_{\rm sig}=\alpha r_{\rm sig}=
  {r_{\rm c}T/2\over1+(r_{\rm bg}+r_2T)\tau_{\rm d}/4}. \label{eqn:rs}
\end{equation}

\noindent For typical parameters in our experiment ($r'_1$=78\,kcps,
$r'_2$=71\,kcps, $r'_{\rm c}$=11\,kcps, $\tau_{\rm d}=1\,\mu$s, $T$=15\%,
$q_{\rm i}$=4.3\%, $\tau_{\rm c}$=2\,ns), the total detector rate $r_{\rm
  t}=\alpha(r_{\rm bg}+r_2T)$ on the receiver side, the available raw key bit
rate $r'_{\rm sig}$ and the resulting QBER $q_{\rm t}$ are plotted as a
function of an external background rate $r_{\rm bg}$ in
figure~\ref{fig:saturation}. Above a certain background rate, $q_{\rm t}$ would
exceed the limit of 11\% for which a secret key can be established for
individual attack schemes \cite{scarani:08}.

It is instructive to consider the excess QBER due to background events:

\begin{equation}
\Delta q=q_{\rm t}-q_{\rm i}=(r_1-Tr_{\rm c})\,r_{\rm bg}\tau_{\rm c}\,{1/2-q_{\rm
    i}\over r_{\rm c}T+r_1r_{\rm bg}\tau_{\rm c}}\,.
\end{equation}

\noindent In a parameter regime useful for key generation, $q_{\rm i}\ll1/2$,
$r_{\rm sig}\gg r_a$, and for simplicity assuming $r_1\gg Tr_{\rm c}$, this
quantity can be approximated by

\begin{equation}
\Delta q\approx{r_{\rm bg}\tau_{\rm c}\over2 T (r_{\rm c}/r_1)}\,. \label{eqn:dq}
\end{equation}

\noindent While the source property $r_{\rm c}/r_1$ and the channel
transmission $T$ are typically optimized already, the only way to reduce the
excess error $\Delta q$ is to reduce the background rate $r_{\rm bg}$ and the
coincidence time window $\tau_{\rm c}$. The limitation on reducing $\tau_{\rm
  c}$ is the timing jitter of all detectors, which in our case is on the order
of a nanosecond. Emphasis thus has to be drawn to reduce the background rate
$r_{\rm bg}$. 

\section{Experimental setup}

\begin{figure}
\begin{center}
\includegraphics{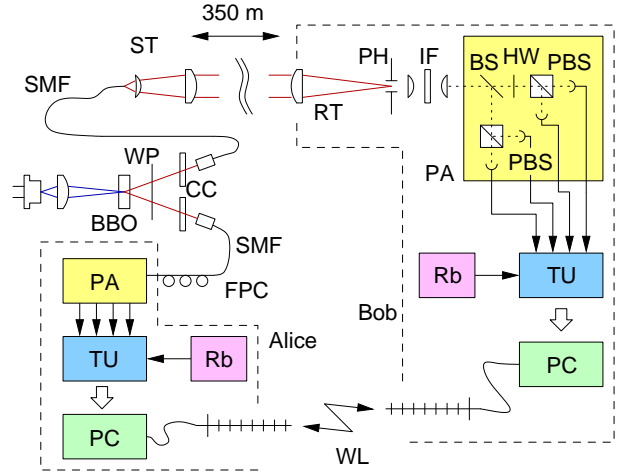} 
\end{center}
\caption{{Schematic diagram of the QKD setup. Components are a sending
    telescope (ST), single mode fibers (SMF), a waveplate (WP), compensating
    crystals (CC) to address birefringent walk-off; polarization analyzer
    units (PA) comprising a 50:50 Beam splitter (BS), polarizing beam
    splitters (PBS) and a half wave plate (HW); a timestamp unit (TU)
    referenced to a Rb oscillator (Rb), a receiving telescope (RT) with a
    pinhole (PH) for spatial filtering and an interference filter (IF) for
    spectral filtering.}} \label{fig:setup}
\end{figure}

We prepare the polarization-entangled photon pairs in a source based
on type-II parametric down conversion (PDC) in a non-collinear configuration
\cite{kwiat:95}. It is pumped with a CW free-running diode laser with power of
30\,mW and a center wavelength of 407\,nm, producing pairs at a degenerate
wavelength around 814~nm
(similar to \cite{trojek:04}) in single mode fibers. When directly
connected to single photon detectors, we typically observe single rate per arm
of 78\,kcps and 71\,kcps, with a coincidence rate of
12\,kcps. The visibility of polarization correlations in the HV and
$\pm45^{\circ}$ basis are $97.5\pm0.5$\% and $92.1\pm0.8$\%,
respectively. While these sources have been substantially surpassed 
in quality and brightness \cite{fedrizzi:07, trojek:08}, this particular
device is both simple and robust.

The minimal incident angle $\gamma$ of the sun and the line of sight was about
$16^\circ$. As endpoints in our transmission channel, we
use a pair of custom telescopes to transmit one member of the entangled photon
pair across a distance of 350\,m for convenient logistics. The relative
orientation of both telescopes is adjusted using manual tip/tilt stages with
an angular resolution of $\approx$10\,$\mu$rad, mounted on tripods intended
for mobile satellite links. The telescopes are not actively stabilized, but
this could be added for spanning larger distances, or to compensate for
thermal drifts in the mounting stages \cite{bienfang:04, weier:06, ursin:07}.

Similarly to \cite{marcikic:06}, the sending telescope consists of
fiber port, a small achromat with $f=100\,$mm to reduce the effective
numerical-aperture of the single mode fiber, and a main achromat with
$f=310$\,mm and 75\,mm diameter, transforming the optical mode of the fiber to
a collimated Gaussian beam with a waist parameter of 20\,mm. Nominally this
results in a Rayleigh length of 1.6\,km at our operation wavelength, well
above our target distance.

A combination of spectral, spatial and temporal filtering is used to reduce
the background to tolerable levels. At the receiving end, an identical
$f=310$\,mm achromat 
as the front lens focuses the incoming light onto a pinhole of 30\,$\mu$m
diameter at its focal position for spatial filtering. Assuming
diffraction-limited performance of that lens, this corresponds to a solid
angle of $2.3\times10^{-9}$\,sr. The pinhole is then imaged with a
magnification of 6.8 through an interference filter onto the passively
quenched silicon avalanche detectors with an active diameter of 500\,$\mu$m in
a compact module that performs passively the random basis selection for the
measurement  \cite{kurtsiefer:02b} (see figure~\ref{fig:setup}).

\begin{figure}
\begin{center}
\includegraphics{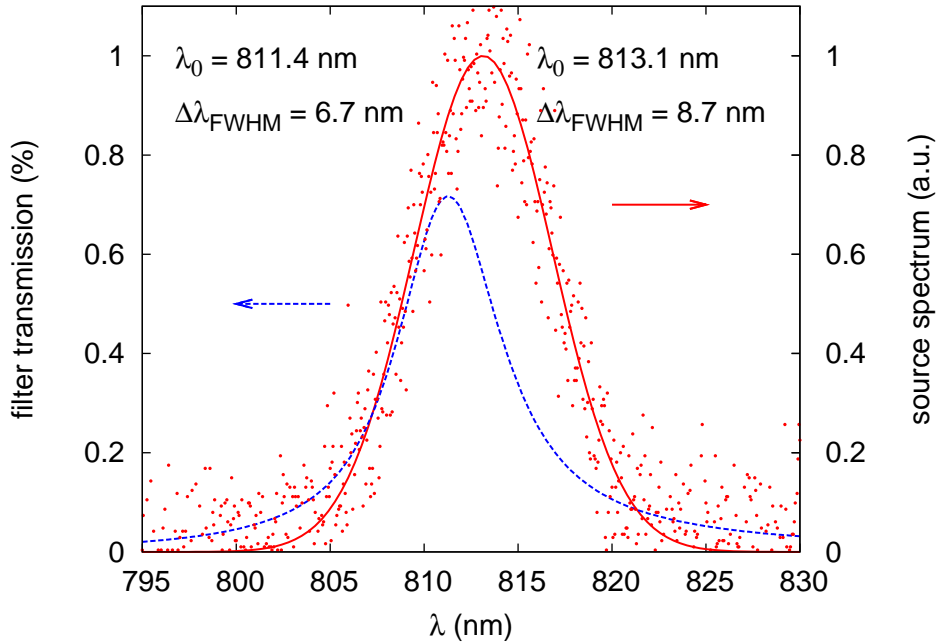} 
\end{center}
\caption{Spectral distribution of photons from the SPDC source, and
  transmission profile of the interference filter used to suppress background
  light outside that range. With this filter/source combination, a signal loss
  of 57\% is introduced.} \label{fig:spectra} 
\end{figure}

 Our pair source has a measured spectral
width of 8.7\,nm, given by the phase matching conditions, and the
geometry of the collection \cite{ling:08b}. An interference filter with a peak
transmission of 72\% and a full width at half-maximum of 6.7\,nm was chosen to
maximize the amount 
of signal transmitted and eliminate the background outside of the spectral
region of the source (see figure~\ref{fig:spectra}). This filter reduces the
ambient background level by about two orders of magnitude, much less than
what can be achieved in PaS experiments based on extremely narrow band lasers
and matching spectral filters.

\begin{figure}
\begin{center}
\includegraphics{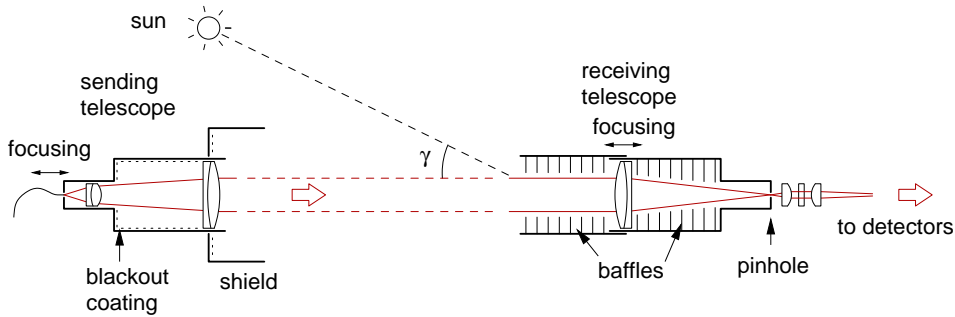} 
\end{center}
\caption{Schematic of the optical ports. To reduce the scattering from ambient
radiation into the optical path, the transmitter telescope areas are shielded
and coated inside with diffuse blackout material. On the receiver side,
several baffles reduce the impact of strong ambient light entering the
telescope under small angles.} \label{fig:telescopes}
\end{figure}

A significant reduction of background events was also achieved addressing
scattering from various elements in the field-of-view (FOV) of the detector,
and from scatterers close to the optical channel (see
figure~\ref{fig:telescopes}). Reduction of the FOV with a smaller pinhole is
finally limited by diffraction; we found a 30\,$\mu$m pinhole to be the optimal
choice when considering pointing accuracy and signal
transmission. This corresponds to a FOV of $\approx73$\,mm diameter for our
test range which will strongly contribute to daylight background counts. 
A circular area with a diameter of about 3 FOV as well as the inside of the
sending telescope is covered with low scattering blackout material. The
blackout area was also shielded against direct sunlight. Together, 
these steps reduce the background by about 12\,dB. 
A set of apertures at the receiver telescope removes light coupled
to the detector by multiple reflections from outside the
line-of-sight. Five concentric apertures extending 30\,cm upstream, and seven
apertures with tapered diameters downstream of the main receiver lens matched
the receiving mode and reduced the background by about 3-4\,dB.

The processing of detection events into a final key has been described in 
\cite{marcikic:06} and the software is available as open source
\cite{googlecode}. 
Each detector event results in a NIM pulse which is sent to a custom
timestamp unit with a nominal resolution of 125\,ps, referenced to a local Rb
oscillator. Our time stamp units exhibit a dead time of 128\,ns, and are able to
transfer up to $6 \cdot 10^6$ events per second to a commodity host PC via a USB
connection. The timing information on one of the sides is then losslessly
encoded as differences between consecutive events with an adaptive resolution,
and together with the basis information sent to the other side on a classical
channel, in our case over a standard wireless TCP/IP
connection. The encoding, together with a small overhead, consumes about 13\%
more bandwidth than necessary due to the Shannon limit. To minimize the
bandwidth for this communication, the timing information was sent from the
source side with lower overall detection rate during daylight conditions.

To identify corresponding photon pair events, the temporal correlation of the
two photons generated in the PDC process is used 
\cite{burnham:70}, with a coincidence time window determined by
the combined timing jitter from both photodetector sets, the timestamp
electronics, and the time difference servoing.

For that process to work, an initial time difference between the two receiver
units due to different timing origins and light propagation is determined to a
resolution of 2\,ns using a tiered cross correlation technique on a set of
detector events acquired over $\approx 5$ seconds. Once that time
difference is established, coincidences are identified within a time window
of $\tau_{\rm c}=2\,$ns. Its center drift due to residual frequency differences
between the two reference clocks is tracked with a servo loop with an
integration time constant of 2\,s for events falling in a time window
$\pm3.75\,$ns around the expected center for coincidence events. We were able
to resynchronize the system during daylight conditions at a coincidence rate
of 1\,500\,cps up to an ambient light level of 250\,kcps, well below saturation of
the detectors.

To maintain a common time frame when no useful signal is available for
servoing, one of the clock
frequencies was manually adjusted such that the relative frequency difference
was $\approx10^{-12}$. This would allow a loss of signal over a period of
two hours without loss of timing lock. Again, the tight time
correlation of the photon pairs emerging in PDC acts as a natural way of
comparing differences and synchronizing clocks at a distance easily. The
ability to resynchronize during daytime and the use of the PDC signal for
mutual calibration of the clocks makes this system very robust against signal
interruptions or temporal unavailability of the channel. 

\begin{figure}
\begin{center}
\includegraphics{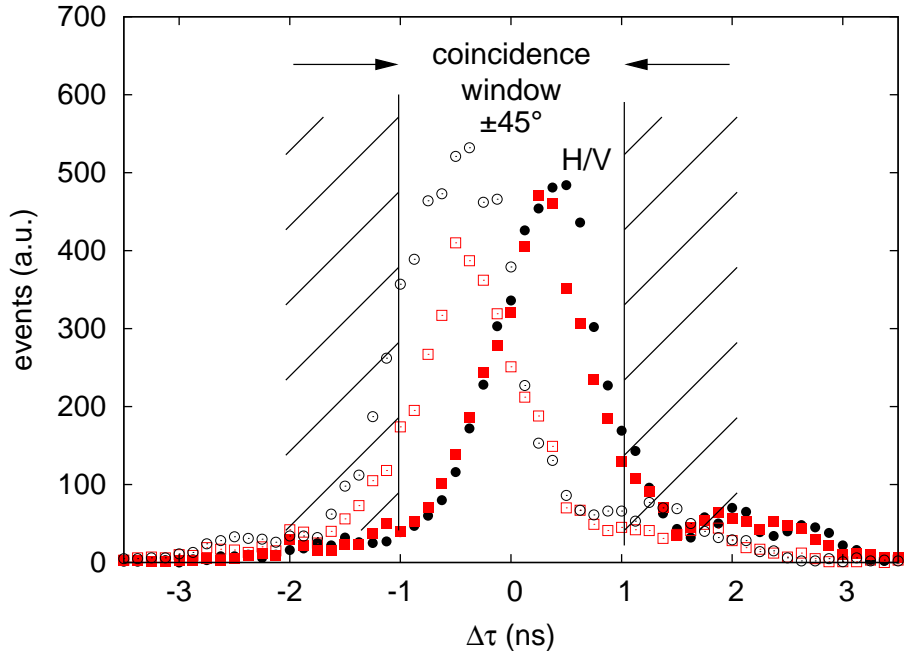} 
\end{center}
\caption{Histograms of time delays between the four main coincidence
  combinations contributing to the raw key. The overlap between detector pairs
  operating in the same basis is excellent, but there is approximately 0.5\,ns
  difference between the two detector groupings. The coincidence window of
  2\,ns is indicated in the graph, showing that we are loosing some key 
  generating counts. If the detector groups were mutually compensated, the
  coincidence window could have be tightened with no loss of signal, but
  reducing the background proportionally.} \label{fig:histos} 
\end{figure}

In the discussion of temporal filtering we have assumed that all detectors on one side have the same relative lag, or more generally, that their temporal response is identical. This detector equivalence is not guaranteed, and is necessary for an efficient time filtering and, more importantly, to prevent information leakage to an eavesdropper \cite{lamas-linares:07a}.
Figure~\ref{fig:histos} shows the measured time differences between those
pairs of detector combinations which contribute to the key generation. The
figure shows that detectors from the same basis are well matched, but there is
significant difference between the two bases. Given our detector assignment,
the information leakage is 0.52\% and 0.44\% in the HV and $\pm 45^\circ$
basis, respectively. For continuously pumped sources, extraction of timing
information by an eavesdropper would need a measurement of the
presence of a photon in the communication channel without disturbing the
polarization state; with pulsed sources, however, the problem becomes more
acute, as the pulse train provides a clock with which to compare the publicly
exchanged timing information. 

\section{Experimental results}
\begin{figure}
\begin{center}
\includegraphics[scale=1]{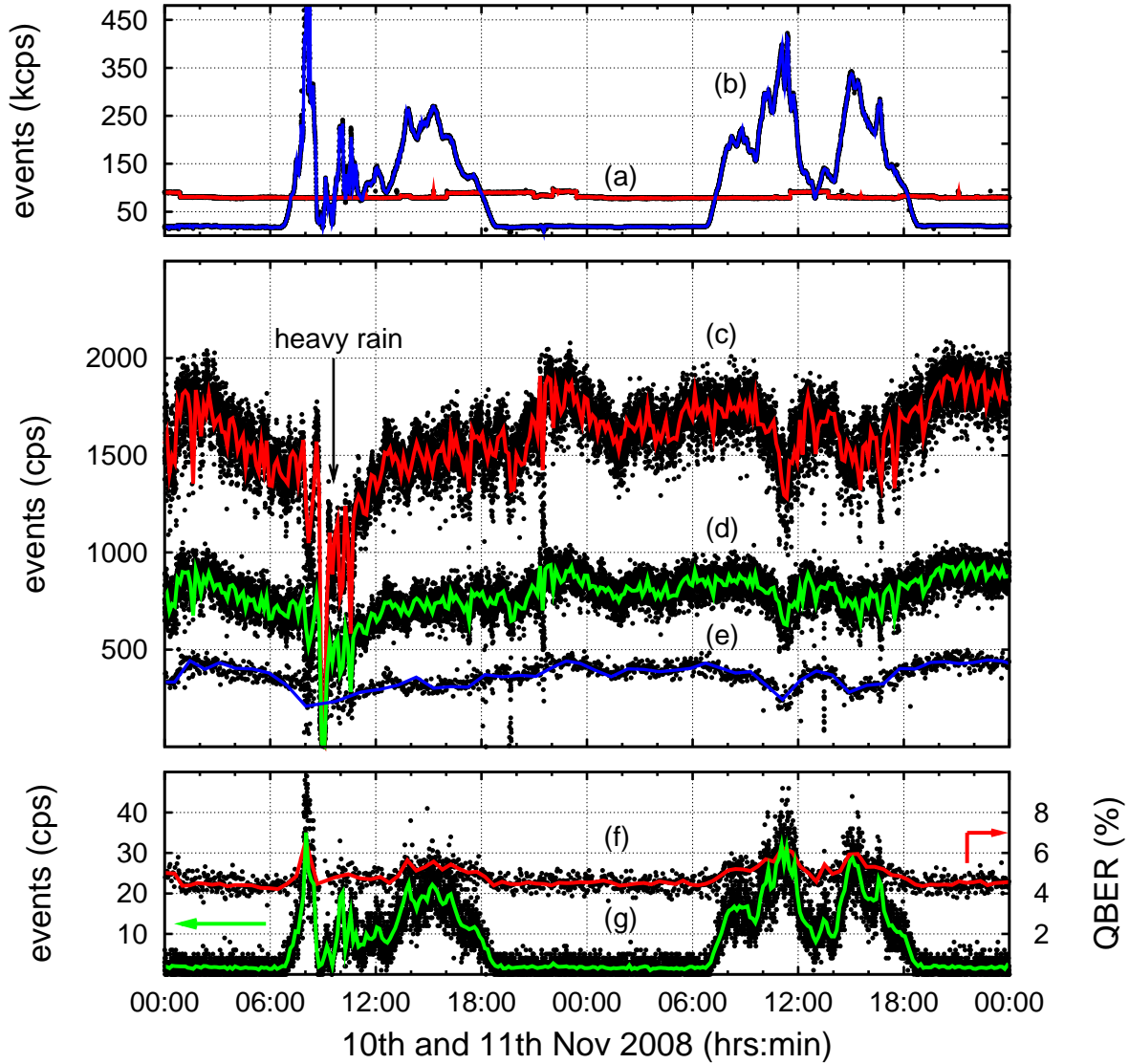} 
\end{center}
\caption{{(color online) Experimental results for 10th and 11th November
    2008. The top panel records the firing rate of the single photon
    detectors. The stable trace (a) corresponds to a detector connected
    directly to one arm of the source and isolated from changes in ambient
    light. The other trace (b) is the detector coupled to the free space
    channel. The middle panel traces show (c) the number of
    raw pair events, (d) sifted events, and (e) error corrected and privacy
    amplified  key. The lower panel shows the number of ``accidental'' pair
    events detected (g), and the QBER level as a percentage
    (f). All experimental points are sampled down, and the solid lines
    represent a moving average as a guide to the eye.}} \label{fig:24hrs}
\end{figure}

The experiment was run continuously over a period from 9.11.2008, 18:00 SGT to
14.11.2008, 2:00 SGT over four consecutive days. In this period we saw
extremely bright sunlight, tropical thunderstorms and partly cloudy 
weather; over the whole period the rate of detected pairs and
background events varied by about 2 orders of
magnitude. In figure~\ref{fig:24hrs} we show the results collected over two
consecutive days. On the second day we identified
$14.72 \cdot 10^7$ raw coincidences. After sifting, this resulted in $7.18
\cdot 10^7$ of raw uncorrected bits, with a total of $3.5 \cdot 10^6$
errors corrected using a modified CASCADE protocol \cite{cascade}, which was
carried out over blocks of at least 5\,000 bits to a target bit error ratio of
$10^{-9}$.

For the privacy amplification step, we arrive at a knowledge of an
eavesdropper on the error-corrected raw key determined by (a) the actual
information revealed in the error correction process, and (b) the asymptotic
(i.e. assuming infinite key length) expression for the eavesdropping knowledge
inferred from the actually observed QBER $q_T$,
$I_E=-q_t\log_2q_t-(1-q_t)\log_2(1-q_t)$ of an equivalent true single photon
BB84 protocol. Privacy amplification itself is carried out by binary
multiplication/addition 
of blocks of raw key vectors with a length of at least 5\,000 bits with a
rectangular matrix filled with a pseudorandom balanced bit stream from a 32 bit
linear-feedback shift register,
seeded with a number from a high-entropy source for each block.
We are left with $3.33\cdot 10^7$ of secure bits for this 24 hour period,
corresponding to an average key generation rate of 385 bits per second (bps). In
these conditions, the key generation rates are far from uniform during the
acquisition period; we see a maximum secure key generation rate of 533\,bps in
darkness and a minimum of 29\,bps around noon in rainy conditions.

The raw key compression ratio in the privacy amplification step should
actually also take care of a limited entropy in the raw key due to
part-to-part variation in detector efficiencies. This information was obtained
before the main key generation process by establishing the complete correlation
matrix (see table \ref{tab1}) out of an ensemble of 148\,493 coincidence
events with matching bases.
\begin{table}
\begin{center}
  \begin{tabular}{c||c|c|c|c}
    Events&H&$+45^{\circ}$&V&$-45^{\circ}$ \\ \hline \hline
    H&599&22\,791&34\,032&18\,409\\ \hline
    $+45^{\circ}$&18\,647&2\,894&17\,512&44\,841\\ \hline
    V&29\,062&16\,422&2\,125&25\,246\\ \hline
    $-45^{\circ}$&14\,635&40\,558&22\,280&1\,498
  \end{tabular}
\end{center}
  \caption{Correlation events between each of the four detectors
    on both sides}\label{tab1}
\end{table}
The asymmetry between 0 and 1 results in the HV basis
is $53.9:46.1\pm0.2\%$, and in the $\pm45^\circ$ basis
$52.5:47.5\pm0.2\%$. Using again entropy as a simple measure of information
leakage, this detector asymmetry would allow an eavesdropper to obtain 0.45\%
of the raw key for events in the HV basis, and 0.18\% in the $\pm45^\circ$
basis. At the moment, however, it is not obvious that a simple reduction
of the final key size in the privacy amplification step due to various
information leakage channels would be sufficient to ensure that the
eavesdropper has no access to any elements of the final key. We also note that
the choice between the two measurement bases is not completely balanced; the
ratio of HV vs. $\pm45^{\circ}$ coincidences is
$42.5:57.5\pm0.1\%$. Furthermore, this asymmetry varies over time. For the
combined asymmetry between 
logical 0 and 1 bits in the raw key we find around 51.5\% during night time, and
54.0\% during daytime. A system which captures this variability in
detection efficiencies (and also would allow to discover selective detector
blinding attacks) would have to monitor this asymmetry continuously.

\begin{figure}
\begin{center}
\includegraphics{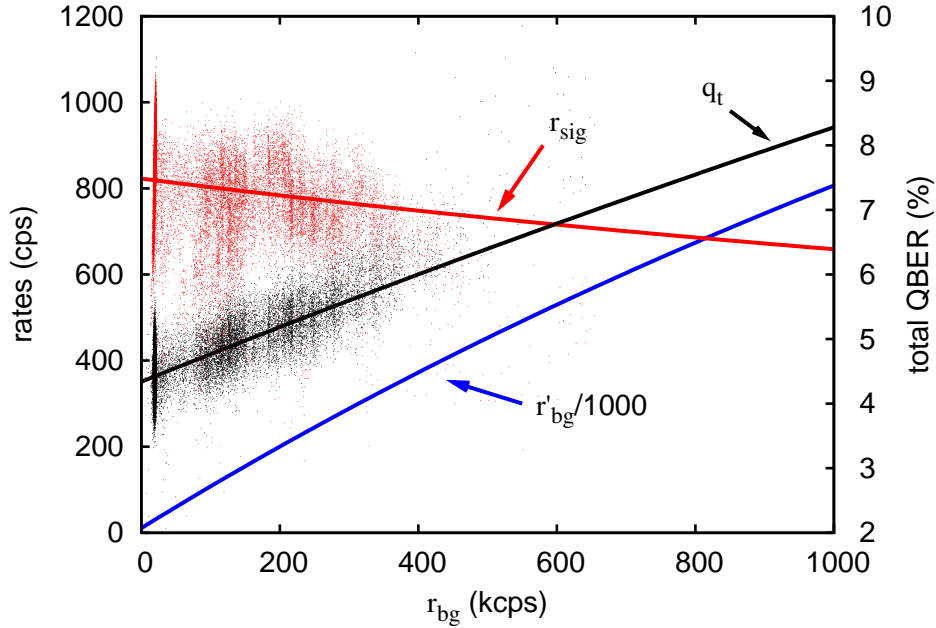} 
\end{center}
\caption{{
Detection behavior due to an external background rate $r_{\rm bg}$ for
parameters representative for our experiment: The \emph{detected} background
rate $r'_{\rm bg}$ shows saturation, due to the intrinsic dead time of the
four detectors. Less counts are detected for higher count rates. The observed
background rate increases up to $\approx$ 450\,kcps, which leads also to a
reduction of the sifted key rate, $r_{\rm sig}$, by 20\% and an increase of
the resulting QBER $q_{\rm t}$ up to 6.5\%. The efficient filtering of the
ambient light prevents a higher background, which would lead to an increase of
the QBER above the threshold of 11\% where no private key can be established
between the two parties at a count rate of 1.8$\cdot 10^6$\,cps. This
threshold is not reached during the whole experiment, thus continuous
operation is possible when the coupling between the parties is
maintained. }}\label{fig:saturation} 
\end{figure}

As introduced in section \ref{chap:theory} we can estimate how well the
experiment performs for a given number of background
events. Figure~\ref{fig:saturation} shows theoretical values for background-
and signal rates according to equations (\ref{eq:qbertot}) and (\ref{eqn:rs}), and
experimental data for the 25\,000 recorded outputs of the error correction
module during the two days of the experiment.  The dead-time affected detector
response is also shown assuming $\tau_d=1\,\mu$s. 
The night time periods with $r_{\rm sig}\approx12$\,kcps and a total dark
count rate of 7\,kcps contribute to events on the low background regime of
the experiment forming a vertical line to the left of
figure~\ref{fig:saturation}. We cannot differentiate between fluctuations due to
changes in the source and those in the transmission channel, but since the
source itself was protected against thermal fluctuations, we attribute them to
variations in transmission $T$ due to changes in the coupling of the 
telescopes. The strongly fluctuating background during daytime
contributes to the broadly scattered data between $r_{\rm bg}$ = 20 and
500\,kcps. If the source properties and channel coupling were constant, the 
deviation of $q_t$ and $r_{\rm sig}$ from the theoretical value would be
both randomly distributed. Figure~\ref{fig:saturation}, however, shows more
structure in $r_{\rm sig}$ than in $q_{\rm t}$ which we attribute to changes
in the coupling between the telescopes due to thermal expansion. Nevertheless,
the experimental values fit the theoretical prediction well. We note that
saturation of the detectors is never a problem.

There are two contributors to the variability in key generation rate: First,
atmospheric conditions such as rainfall reduce the transmission and thus
the number of raw key events before the error correction and privacy
amplification steps, but the QBER remains unchanged. On the
other hand we have extremely bright conditions where accidental
coincidences increase significantly. In this regime, as the background rises,
the signal rate is reduced due to the dead time of the detectors. Furthermore,
the QBER increases according to equation~(\ref{eq:qbertot}), 
occasionally preventing the generation of a secure key \cite{scarani:08}. But 
 even under bright conditions, the system still keeps track of the time drift
 between the two reference clocks with the time-correlated coincidences from
 the source without a need for re-synchronization. 

\section{Conclusions}
We have demonstrated the continuous running of a free space entanglement QKD
system over several full day-night cycles in variable weather conditions. A
combination of filtering 
techniques is used to overcome the highly variable illumination and
transmission conditions. The software and synchronization scheme can tolerate
the remaining 16\,dB variation in light levels without interruption of the key
generation. We continuously generate error corrected, privacy amplified key at
an average rate of 385\,bps. 
With the newly available bright sources larger distances and/or a higher key
generation rates are possible. 

\section*{References}

\end{document}